# Discovery of Powerful Gamma-Ray Flares from the Crab Nebula

M. Tavani[1,2,3,4], A. Bulgarelli[5], V. Vittorini[1], A. Pellizzoni[16], E. Striani[2,4], P. Caraveo[7], M.C. Weisskopf[21], A. Tennant[21], G. Pucella[6], A. Trois[1], E. Costa[1], Y. Evangelista[1], C. Pittori[19], F. Verrecchia[19], E. Del Monte[1], R. Campana[1], M. Pilia[16,17], A. De Luca[7,25], I. Donnarumma[1], D. Horns[22], C. Ferrigno[23], C.O. Heinke[24], M. Trifoglio[5], F. Gianotti[5], S. Vercellone[18], A. Argan[1], G. Barbiellini[3,8,9], P.W. Cattaneo[10], A.W. Chen[3,7], T. Contessi[7], F. D'Ammando[18], G. DeParis[1], G. Di Cocco[5], G. Di Persio[1], M. Feroci[1], A. Ferrari[3,11], M. Galli[12], A. Giuliani[7], M. Giusti[1,3], C. Labanti[5], I. Lapshov[13], F. Lazzarotto[1], P. Lipari[14,15], F. Longo[8,9], F. Fuschino[5], M. Marisaldi[5], S. Mereghetti[7], E. Morelli[5], E. Moretti[8,9], A. Morselli[4], L. Pacciani[1], F. Perotti[7], G. Piano[1,4], P. Picozza[1,4], M. Prest[17], M. Rapisarda[6], A. Rappoldi[10], A. Rubini[1], S. Sabatini[1,4], P. Soffitta[1], E. Vallazza[9], A. Zambra[3,7], D. Zanello[14,15], F. Lucarelli[19], P. Santolamazza[19], P. Giommi[19], L. Salotti[20], G.F. Bignami[25].

1. *INAF-IASF Roma, via del Fosso del Cavaliere 100, 00133 Roma, Italy*
2. *Dipartimento di Fisica, Università degli Studi di Roma "Tor Vergata", via della Ricerca Scientifica 1, 00133 Roma, Italy*
3. *Consorzio Interuniversitario Fisica Spaziale (CIFS), villa Gualino, v.le Settimio Severo 63, 10133 Torino, Italy*
4. *INFN Roma Tor Vergata, via della Ricerca Scientifica 1, 00133 Roma, Italy*
5. *INAF-IASF Bologna, via Gobetti 101, 40129 Bologna, Italy*
6. *ENEA Frascati, via Enrico Fermi 45, 00044 Frascati(RM), Italy*
7. *INAF-IASF Milano, via E. Bassini 15, 20133 Milano, Italy*
8. *Dip. Fisica, Università di Trieste, via A. Valerio 2, 34127 Trieste, Italy*
9. *INFN Trieste, Padriciano 99, 34012 Trieste, Italy*
10. *INFN Pavia, via Bassi 6, 27100 Pavia, Italy*
11. *Dipartimento di Fisica Generale, Università degli Studi di Torino, via P. Giuria 1, 10125 Torino, Italy*
12. *ENEA Bologna, via don Fiammelli 2, 40128 Bologna, Italy*
13. *IKI, Moscow, Russia*
14. *INFN Roma 1, p.le Aldo Moro 2, 00185 Roma, Italy*
15. *Dipartimento di Fisica, Università degli Studi di Roma "La Sapienza", p.le Aldo Moro 2, 00185 Roma, Italy*
16. *INAF Osservatorio Astron. Cagliari, Poggio dei Pini, 09012 Capoterra, Italy*
17. *Dipartimento di Fisica, Università degli Studi dell' Insubria, via Valleggio 11, 22100, Como, Italy*
18. *INAF IASF Palermo, via La Malfa 153, 90146 Palermo, Italy*
19. *ASI Science Data Center, ESRIN, 00044 Frascati, Italy*
20. *Agenzia Spaziale Italiana, viale Liegi 26, Roma, Italy*
21. *NASA, Marshall Flight Space Center, Huntsville, Alabama*
22. *Institut fuer Experimentalphysik, University of Hamburg, Germany*
23. *ISDC, University of Geneva, Switzerland*
24. *University of Alberta, Edmonton, AB T6G 2G7, Canada*
25. *IUSS, Pavia, Italy*

# Abstract

The well known Crab Nebula is at the center of the SN1054 supernova remnant. It consists of a rotationally-powered pulsar interacting with a surrounding nebula through a relativistic particle wind. The emissions originating from the pulsar and nebula have been considered to be essentially stable. Here we report the detection of strong gamma-ray (100 MeV-10 GeV) flares observed by the AGILE satellite in September, 2010 and October, 2007. In both cases, the unpulsed flux increased by a factor of 3 compared to the non-flaring flux. The flare luminosity and short timescale favor an origin near the pulsar, and we discuss *Chandra* Observatory X-ray and *HST* optical follow-up observations of the nebula. Our observations challenge standard models of nebular emission and require power-law acceleration by shock-driven plasma wave turbulence within a ~1-day timescale.

The Crab Nebula (*1*) is a relic of a stellar explosion recorded by Chinese astronomers in 1054 C.E.. It is located at a distance of 2 kpc from Earth, and is energized by a powerful pulsar of spindown luminosity $L_{PSR} = 5\cdot10^{38}$ erg $s^{-1}$, and spin period $P = 33$ ms *(2,3,4 )*. Optical and X-ray images of the inner nebula show (*1, 5, 6, 7*) features such as wisps (composing a torus-shaped structure), knots and the anvil [positioned along the South-East "jet" originating from the pulsar, and aligned with its rotation axis (*6*)]. Wisps, some of the knots, and the anvil are known to brighten and fade over weeks or months (*6, 8*). The Crab Nebula X-ray continuum and gamma-rays up to ~100 MeV energies are modelled by synchrotron radiation, and emission from GeV to TeV energies as inverse Compton radiation by accelerated electrons scattering CMB and nebular photons (*9, 10, 11, 12*).

The AGILE satellite (*13*) observed the Crab Nebula several times both in pointing mode from mid-2007/mid-2009, and in spinning mode starting in November 2009 (see Supporting Online Material, SOM). The AGILE instrument (*13*) monitors cosmic sources in the energy ranges 100 MeV – 10 GeV (hereafter, GeV gamma-rays) and 18 - 60 keV with good sensitivity and angular resolution. With the exception of a remarkable episode in October, 2007 (see below) we obtain, during standard non-active states, an average (pulsar +nebula) flux value (*14*) of $F_g = (2.2 \pm 0.1)\cdot10^{-6}$ ph. $cm^{-2}$ $s^{-1}$ in the range 100 MeV - 5 GeV, for an average photon index $\alpha = 2.13 \pm 0.07$.

During routine monitoring in spinning mode in September, 2010, a strong and unexpected gamma-ray flare from the direction of the Crab Nebula was discovered (*15*) by AGILE above 100 MeV. The flare reached its peak during 19-21 September 2010 with a 2-day flux of $F_{g,p1}= (7.2 \pm 1.4)\cdot10^{-6}$ ph $cm^{-2}$ $s^{-1}$ ($\alpha = 2.03 \pm 0.18$) for a 4.8 s.d. detection above the average flux. It subsequently decayed within 2-3 days to normal average values (Fig. 1, top panel). This flare was independently confirmed by *Fermi*-LAT (*16,17*), and different groups obtained multifrequency data in the following days (*18*). Recognizing the importance of this event was facilitated by a previous AGILE detection with similar characteristics.

AGILE detected indeed another remarkable flare from the Crab in October, 2007 [see also (*14*)]. The flare extended for ~2 weeks and showed an interesting time sub-structure (Fig. 1, bottom panel). The peak flux was reached on 7 October 2007 and the 1-day integration value was $F_{g,p2} = (8.9 \pm 1.1) \cdot 10^{-6}$ ph. $cm^{-2}$ $s^{-1}$ ($\alpha = 2.05 \pm 0.13$) for a 6.2 s.d. detection above the standard flux.

For both the October 2007 and September 2010 events there was no sign of variation of the pulsar gamma-ray signal (*19*,*20*,*21*) during and after these flares, as independently confirmed for the Sept.-2010 event by gamma-ray (*22*), radio (*23*), and X-ray analyses (see SOM). We thus attribute both flares to unpulsed relativistic shock emission originating in the nebula.

In the following, we focus on the September 2010 flare. Optical and X-ray imaging (*18*) shows no additional source in the Crab region during and after the flare. We note that the flaring GeV spectrum is substantially harder than the standard nebular emission (*10*, *11*, *12*). Fig. 2 shows the high-resolution (arcsecond) optical and X-ray images of the nebula obtained 1-2 weeks after the flare by the *Chandra* Telescope and the *Hubble Space Telescope* (HST). A few nebular brightened features are noticeable in both images. The first one is the optical and X-ray anvil feature close to the base of the pulsar jet, a primary site of shocked particle acceleration in the inner nebula (*6*, *8*). Another brightened feature is at a larger distance from the pulsar, and appears as an elongated striation in both the HST and *Chandra* images.

Important constraints can be derived from the gamma-ray flare luminosity and timescale. The peak isotropic gamma-ray luminosity $L_p \approx 5 \cdot 10^{35}$ erg $s^{-1}$ implies for, e.g., a (3-5) % radiation efficiency (*24*,*25*, *26*), that about (2-3) % of the total spindown pulsar luminosity was dissipated at the flaring site. This large value suggests that the production region was close to the pulsar. Also the flare risetime (~1 day) favors a compact emission region of size $L \leq 10^{16}$ cm. The anvil feature is an excellent flare site candidate, also because of its alignment with the relativistic pulsar jet (*1*, *6*, *8*). This region is expected to be dominated by the leptonic current from the polar jets (*24*, *26*).

Gamma-ray flaring from the Crab Nebula provides a unique opportunity to constrain particle acceleration and radiative processes in a nebular environment. Synchrotron emission from a fresh population of shock accelerated electrons/positrons along the pulsar polar jet can explain the flaring emission in the range 0.1-10 GeV. Fig. 3 shows our flare spectral data and two examples of modelling for different assumptions on the particle populations downstream of the shock (a pure electron-positron relativistic Maxwellian distribution, and a distribution modified by a power-law component). Maxwellian and power-law models predict similar synchrotron radiation fluxes in the GeV band as shown in Fig. 3. However, if the emission from the anvil feature is related to the gamma-ray flaring, power-law models can explain also the X-ray emission from that region. Fast cooling of the highest energy particles drastically decreases the GeV flux within a few days, as observed in both the September 2010 and October 2007 flares.

These gamma-ray flares test and constrain theoretical models applicable to pure pair plasmas (*25*, *27*, *28*) or to distributions modified by the presence of ions that resonantly accelerate pairs by magnetosonic waves (*24*, *26*, *29*). The acceleration rate resulting from local wave absorption at the relativistic (electron or ion) cyclotron frequency and from hydrodynamical constraints is determined to be $R_{acc} \sim (\mathrm{day})^{-1}$, implying a flare region size $L \approx 10^{16}$ cm for a standard downstream sound speed. Furthermore, reconciling the synchrotron cooling timescale $\tau \approx (8 \cdot 10^{8}\ \mathrm{sec})\ B^{-2}\ \gamma^{-1}$ (where the magnetic field $B$ is in Gauss, and $\gamma$ is the particle Lorentz factor) with our observations implies, for a Lorentz factor $\gamma \approx (1\text{-}3) \cdot 10^{9}$ of electrons irradiating in the GeV range, a local magnetic field $B \approx 10^{-3}$ G that is 3-10 times the nebular average (*6*, *12*). Both the 2007 and 2010 gamma-ray flares have similar spectral characteristics (Fig. 3). This observation suggests that a common acceleration process produced electron/positron energy distributions with similar physical parameters.

Considering the AGILE exposure of the Crab Nebula, we estimate that 1-2 strong gamma-ray flares actually occur per year. The Crab Nebula is thus not a standard candle at gamma-ray energies. Significant variations of the Crab Nebula high-energy flux have also been recently reported at X-ray *(30)* and TeV *(31)* energies. It remains to be established whether the gamma-ray flares that we report can be attributed to pulsar activity injecting fresh particles in the surroundings, or to major plasma wave instabilities in the nebular environment.

32. We thank the *Chandra* Observatory Director H. Tananbaum, the *Hubble Space Telescope* Director M. Mountain, N. Gehrels and the *Swift* team for their prompt response in carrying out the observations reported in this paper. Research partially supported by the ASI grant no. I/089/06/2.

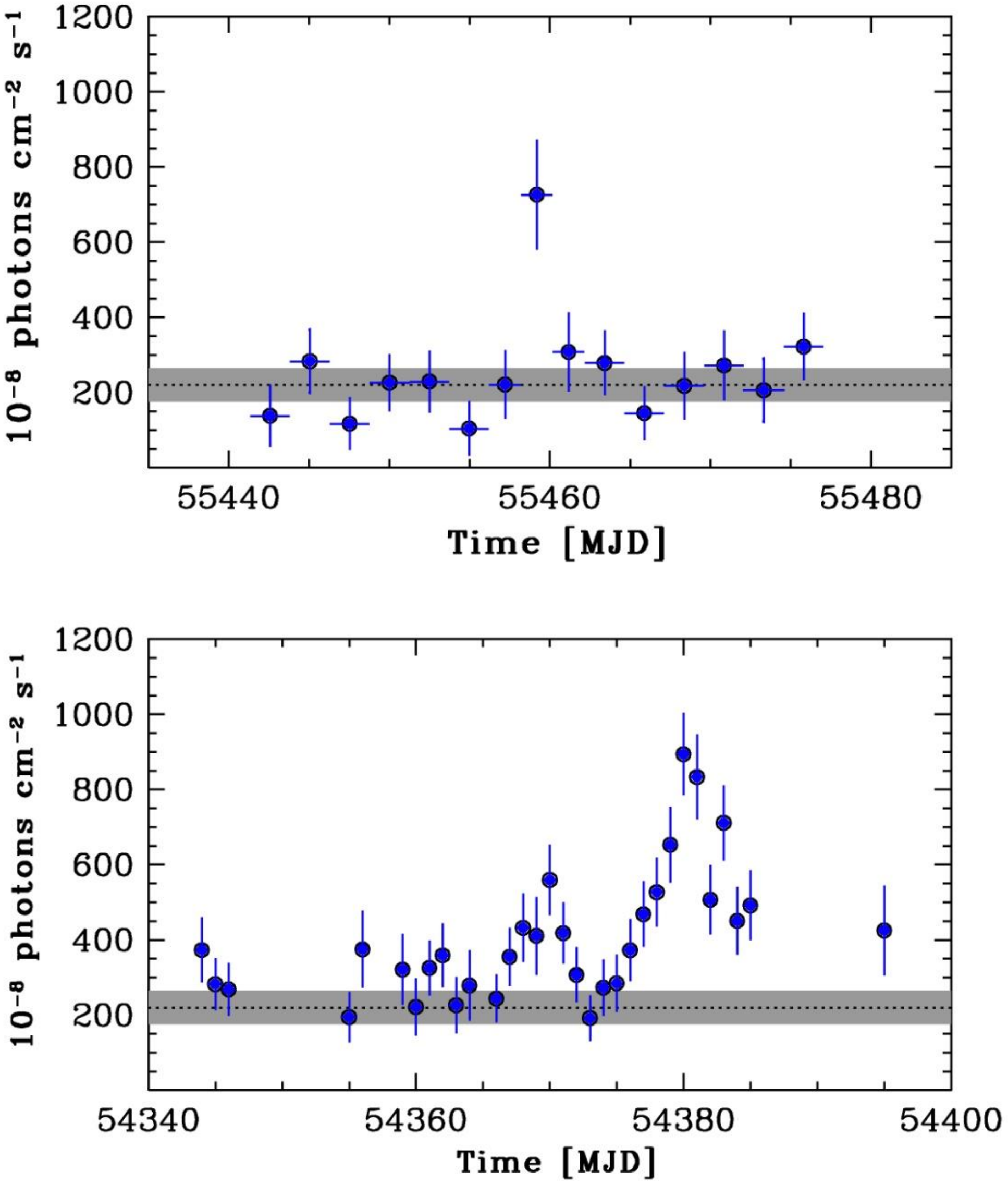

**Fig. 1** – Crab Nebula lightcurves of the total flux detected by AGILE in the energy range 100 MeV – 5 GeV during the gamma-ray flaring periods in 2007 and 2010 (units of $10^{-8}$ ph cm$^{-2}$ s$^{-1}$). *(Top panel:)* the "spinning" AGILE photon flux lightcurve during the period Sept. 2 - Oct. 8, 2010. Time bins are 2.5 days except near the flare peak (2-day binning). Errors are 1 s.d. and time is given in Modified Julian Day (MJD). The dotted line and band marked in grey color show the average Crab flux and the 3 s.d. uncertainty range. *(Bottom panel:)* The AGILE lightcurve during the period Sept. 27 – Oct. 12, 2007 (1-day binning) with the satellite in pointing mode. Errors are 1 s.d. Time is given in Modified Julian Day (MJD). The dotted line and band marked in grey color show the average Crab flux and the 3 s.d. uncertainty range.

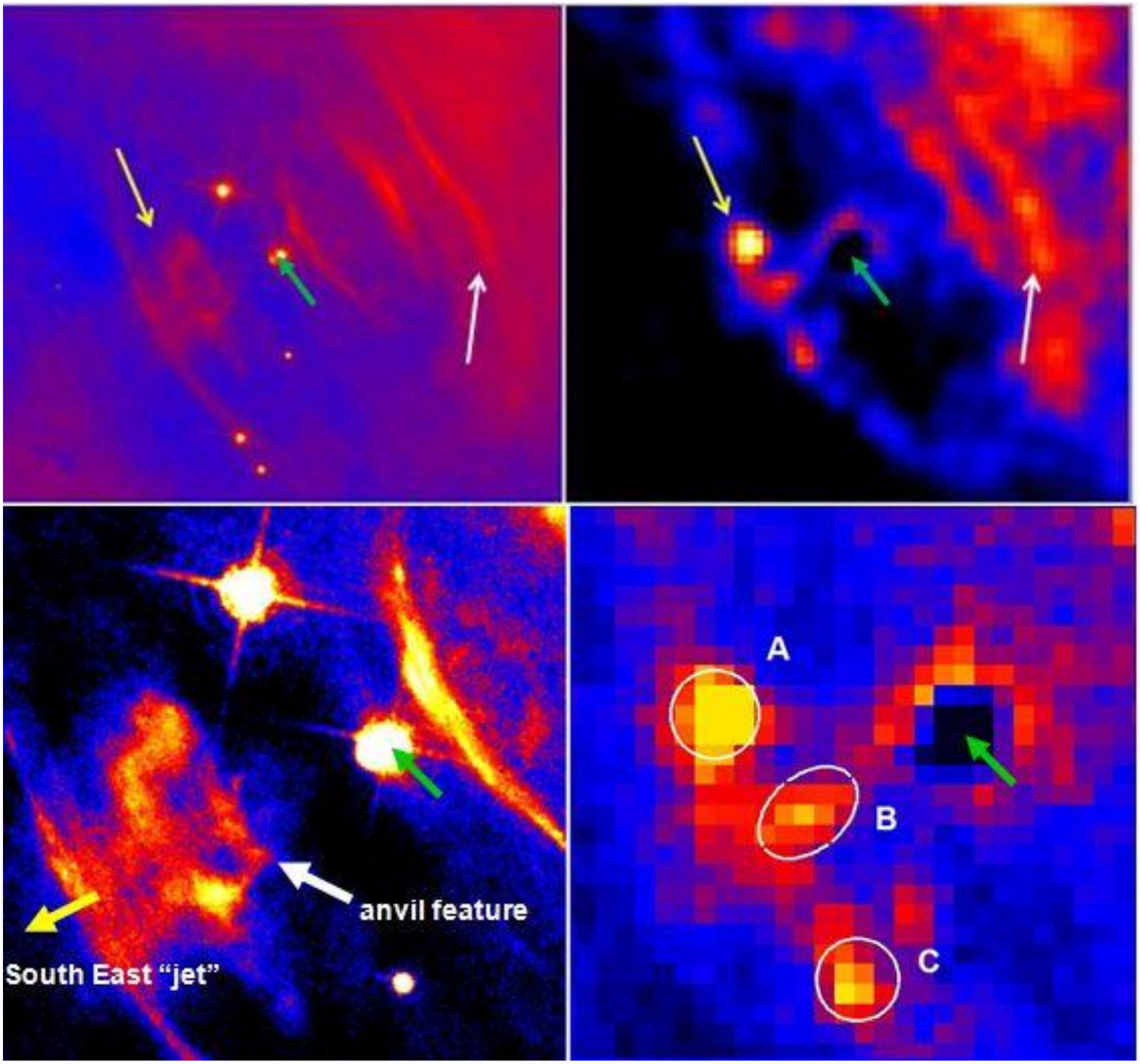

**Fig. 2** – HST and Chandra imaging of the Crab Nebula following the Sept., 2010 gamma-ray flare. *(Top left panel:)* optical image of the inner nebula region (approximately 28"x28", North is up, East on the left) obtained by the ACS instrument on board the *Hubble Space Telescope* (HST) on October 2, 2010. ACS bandpass: 3,500-11,000 Angstrom. The pulsar position is marked with a green arrow in all panels. White arrows in all panels mark interesting features compared to archival data. *(Top right panel:)* the same region imaged by the *Chandra* Observatory ACIS instrument on September 28, 2010 in the energy range 0.5-8 keV (level-1 data). The pulsar does not show in this map and below because of pileup. *(Bottom left panel:)* zoom of the HST image (approximately 9"x9"), showing the nebular inner region, and the details of the "anvil feature" showing a "ring"-like structure at the base of the South-East "jet" off the pulsar. "Knot 1" at 0".6 South-East from the pulsar is saturated at the pulsar position. Terminology is from ref. *6*. *(Bottom right panel:)* zoom of the *Chandra* image, showing the X-ray brightening of the "anvil" region and the correspondence with the optical image. Analysis of the features

marked “A”, “B”, and “C” gives the following results in the energy range 0.5-8 keV for the flux F, spectral index α, and absorption $N_H$ (quoted errors are statistical at the 68% c.l.).

Feature A: flux $F = (48.5 \pm 8.7)\cdot 10^{-12}$ erg $cm^{-2}$ $s^{-1}$, $\alpha = 1.76 \pm 0.30$, $N_H = (0.36 \pm 0.05)\cdot 10^{22}$ atoms $cm^{-2}$.

Feature B: flux $F = (26.6 \pm 5.9)\cdot 10^{-12}$ erg $cm^{-2}$ $s^{-1}$, $\alpha = 1.76 \pm 0.41$, $N_H = (0.34 \pm 0.05)\cdot 10^{22}$ atoms $cm^{-2}$.

Feature C: flux $F = (25.3 \pm 5.9)\cdot 10^{-12}$ erg $cm^{-2}$ $s^{-1}$, $\alpha = 1.46 \pm 0.36$, $N_H = (0.34 \pm 0.04)\cdot 10^{22}$ atoms $cm^{-2}$.

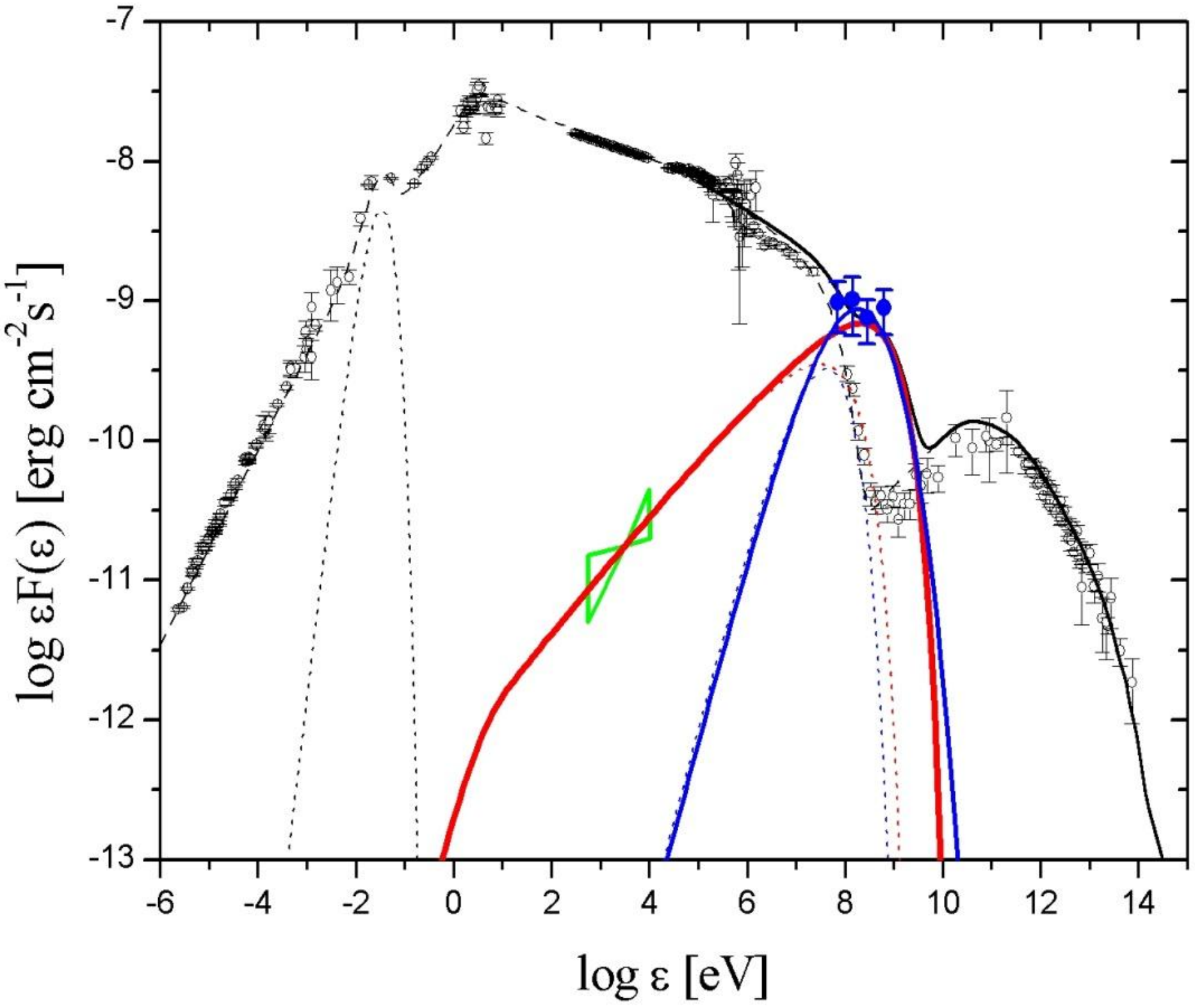

log εF(ε) [erg cm^-2 s^-1]
log ε [eV]

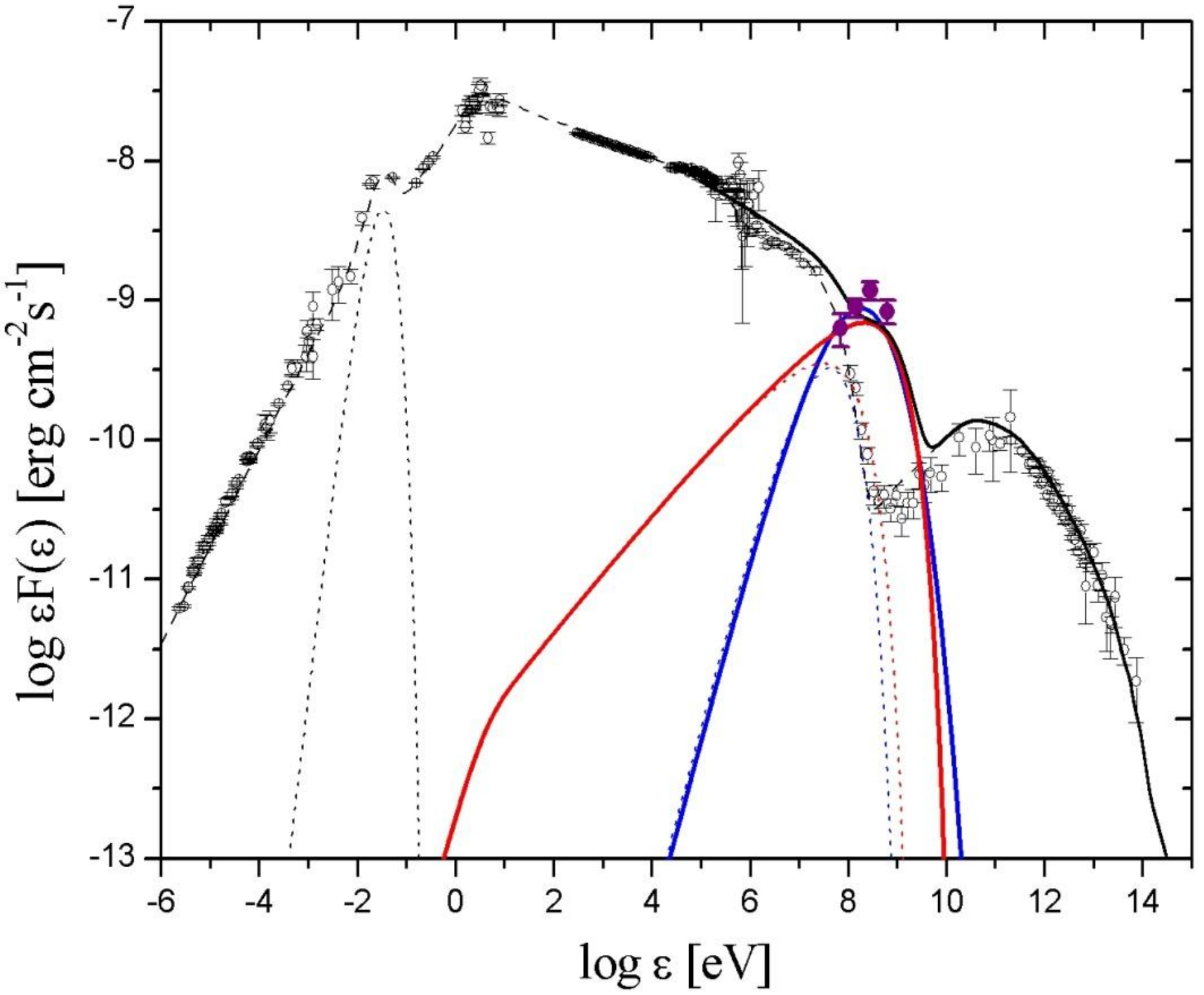

log εF(ε) [erg cm^-2 s^-1]
log ε [eV]

**Fig. 3 -** Spectral energy distribution of the Crab Nebula and the flaring gamma-ray episodes (the pulsar signal has been subtracted). Open black symbols: Crab Nebula emission in the steady state. The dashed curve shows our modelling of the steady state [see also (*1*, *21*, *12*)]. The solid black curve shows our flare modelling for energies above $10^5$ eV. Dotted black curve: nebular IR emission (*1*). (*Top panel:*) blue filled symbols: spectral AGILE gamma-ray flare data integrated over 2 days (September 19-21, 2010, MJD 55458.5-55460.5). Errors are 1 s.d. Solid red and blue curves show the 2-day averaged spectral models based on synchrotron radiation from relativistic electrons/positrons impulsively accelerated in a shock region of size $L \leq 10^{16}$ cm . Dotted curves show the spectra evolved by synchrotron cooling 3 days after the flare. The blue curve model is based on a relativistic Maxwellian distribution of critical energy (Lorentz factor) $\gamma^* = 10^9$ (for a local $B = 10^{-3}$ G) representing the differential energy distribution of accelerated electrons. The red curve model is characterized (for a local $B = 10^{-3}$ G) by a double power-law electron differential distribution $dN(\gamma)/d\gamma = \gamma^{-p1}$ for $\gamma_{min} < \gamma < \gamma_{break}$ with $p_1 = 2.1$, $\gamma_{min} = 5 \cdot 10^5$, $\gamma_{break} = 10^9$, and $dN(\gamma)/d\gamma = \gamma^{-p2}$ for $\gamma_{break} < \gamma < \gamma_{max}$, with $p_2 = 2.7$, $\gamma_{max} = 7 \cdot 10^9$, and a total particle number $N_{e-/e+} = 10^{42}$. This model is extended towards the low-energy range, and can account for the local X-ray spectral enhancements in the "anvil" region as observed by *Chandra* (see Fig. 2). The area marked in green represents the 1 s.d. X-ray spectral data for feature "A" of Fig. 2. (*Bottom panel:*) violet symbols: spectral AGILE gamma-ray data during the October 7-9, 2007 flare (MJD 54380.5-54382.5). Errors are 1 s.d. The black, blue and red curves are those of the Sept. 2010 flare, and are shown here as a reference to compare the two spectra.

# SUPPORTING ON-LINE MATERIAL (SOM)

## Paper entitled: *Discovery of Powerful Gamma-Ray Flares from the Crab Nebula*, by M.Tavani et al.

In this section we provide additional information about the analysis of the AGILE gamma-ray data of the Crab Nebula/Pulsar and of its flaring activity. We also report about the timing analysis of the soft X-ray and hard X-ray data collected by Swift/XRT and SuperAGILE respectively.

### 1. THE AGILE MISSION

AGILE (Astrorivelatore Gamma ad Immagini LEggero - Light Imager for Gamma-ray Astrophysics) is a scientific mission of the Italian Space Agency (ASI) launched on April 23, 2007[1]. The AGILE scientific payload is made of two co-aligned instruments: (1) a gamma-ray imager (GRID) made of a Tungsten-Silicon Tracker[2,3,4] and of a CsI(Tl) Mini-Calorimeter (MCAL) detector[5], with a large field of view (about 2.5 sr), a good time (~100 μs) and excellent angular resolution; (2) a Silicon based imaging hard X-ray detector (named SuperAGILE[6]) with ~1 sr field of view, arcminute angular resolution, dead time of ~5 μs and a typical flux sensitivity of about 15 mCrab (5σ, ~50 ks net exposure time) in the 18-60 keV energy band. The AGILE/GRID instrument has a good sensitivity (the on-axis effective area is about 400 $cm^2$ at 100 MeV), and a source location accuracy (SLA) which can reach values of 0.1-0.2 degrees in the 100 MeV-10 GeV energy range for deep integrations, and 0.5-1 degrees for 1-2 day timescale integrations.

The whole instrument is surrounded by an anti-coincidence (AC) system[7] of plastic scintillators for the rejection of background charged particles. An effective background rejection, event trigger logic, and on-board data storage and transmission is implemented[8]. AGILE orbital characteristics (quasi-equatorial with a inclination angle of 2.5 degrees and average 530 km altitude) are optimal for low-background gamma-ray observations.

### 2. AGILE OBSERVATIONS OF THE CRAB NEBULA/PULSAR

The Crab Nebula is the remnant of a supernova explosion occurred in 1054 AD, observed at the time by Chinese astronomers[9]. The nebula was then re-discovered by John Bevis and Charles Messier in the XVIII century. In 1949 the Crab was identified with the newly discovered very intense radio source Taurus A, and its emission was interpreted[10] as due to the synchrotron emission of high-energy electrons. In 1968 a pulsar approximately coincident with the center of the nebula was discovered in the radio band[11], and soon thereafter also in the optical[12], X-ray[13] and gamma-ray bands[14]. Both the nebula and the pulsar shine in all the bands of the electromagnetic spectrum, from the radio to the very high energy gamma-ray regions. In the X-rays, the Crab Nebula is one of the brightest sources in the sky, and was usually regarded as a stable source[15] thus being an excellent candidate for a "standard candle" for the calibration of X-ray and gamma-ray instruments[16].

Starting from July, 2007 (beginning of the scientific operations) until October, 2009, AGILE operated in fixed-pointing mode, collecting several observations of the Crab Nebula/Pulsar corresponding to a total exposure of ~ $1.2 \cdot 10^9$ $cm^2$ s. Since November, 2009 AGILE is operating in spinning-mode, with the satellite axis sweeping an entire circle in the sky in approximately 7 minutes. Depending on the season, the whole sky is progressively exposed with a typical accumulating pattern. The current total exposure on the Crab Nebula/Pulsar collected by AGILE in spinning-mode is about ~ $5.1 \cdot 10^8$ $cm^2$ s. The signal to noise ratio achieved with a one-day integration in pointing operative mode is roughly reached in two days in spinning mode. For this reason the bin size of the Crab Nebula lightcurves are 1-day in pointing mode (Fig. S1 and S2) and 2-days in spinning mode (Fig. S3 and Fig. 1 of the main text) respectively.

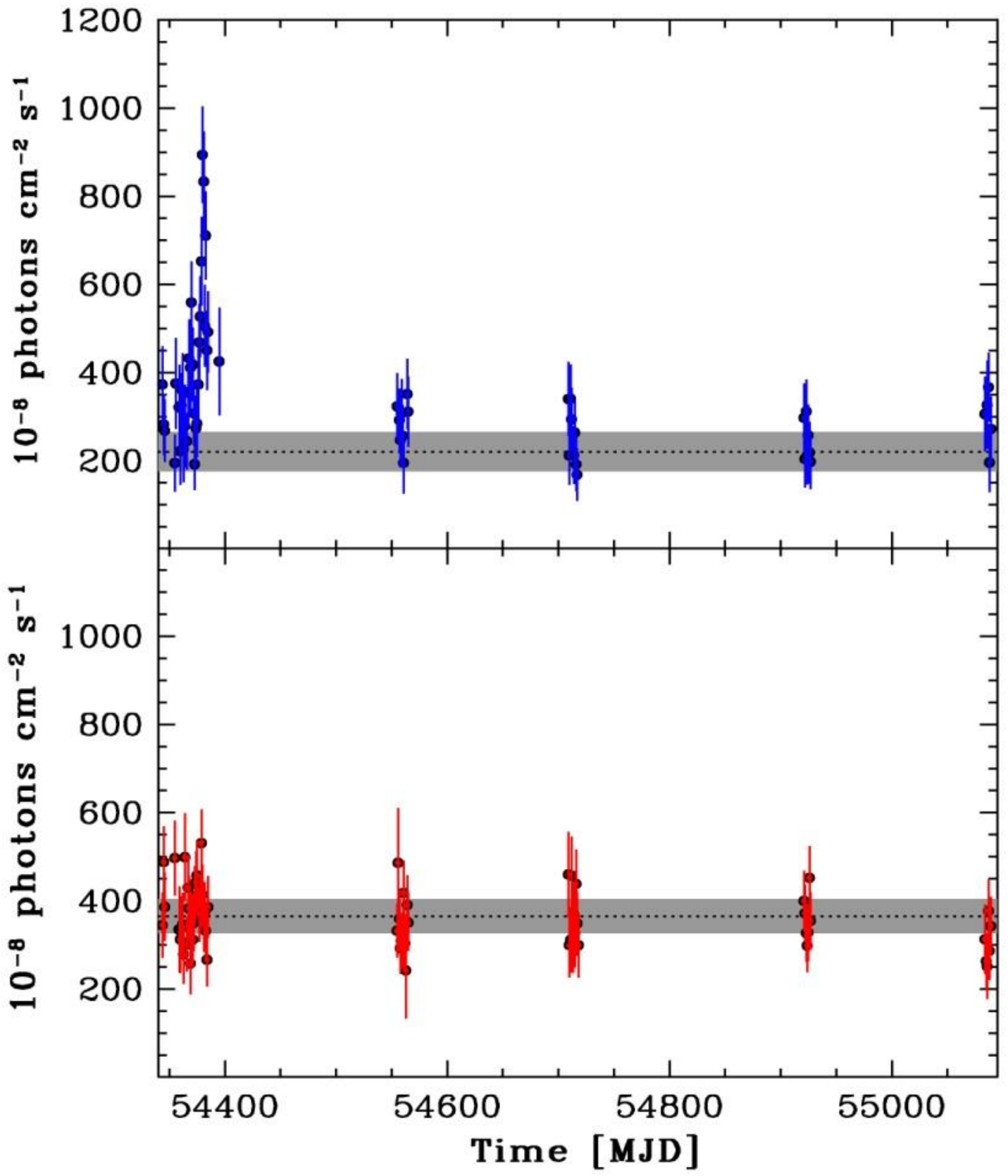

**Fig. S1** – Top panel: The AGILE gamma-ray light curve (1-day binning) of the Crab Pulsar/Nebula above 100 MeV during the period 2007-09-01 – 2009-09-15 with the satellite pointing within 35 degrees from the source. Gaps in the light curve are due to the satellite pointing at fields different from the Crab region. Bottom panel: same as the top panel light curve but for the nearby Geminga pulsar. Dashed lines and shadowed bands indicate the Crab average flux and the 3σ uncertainty range.

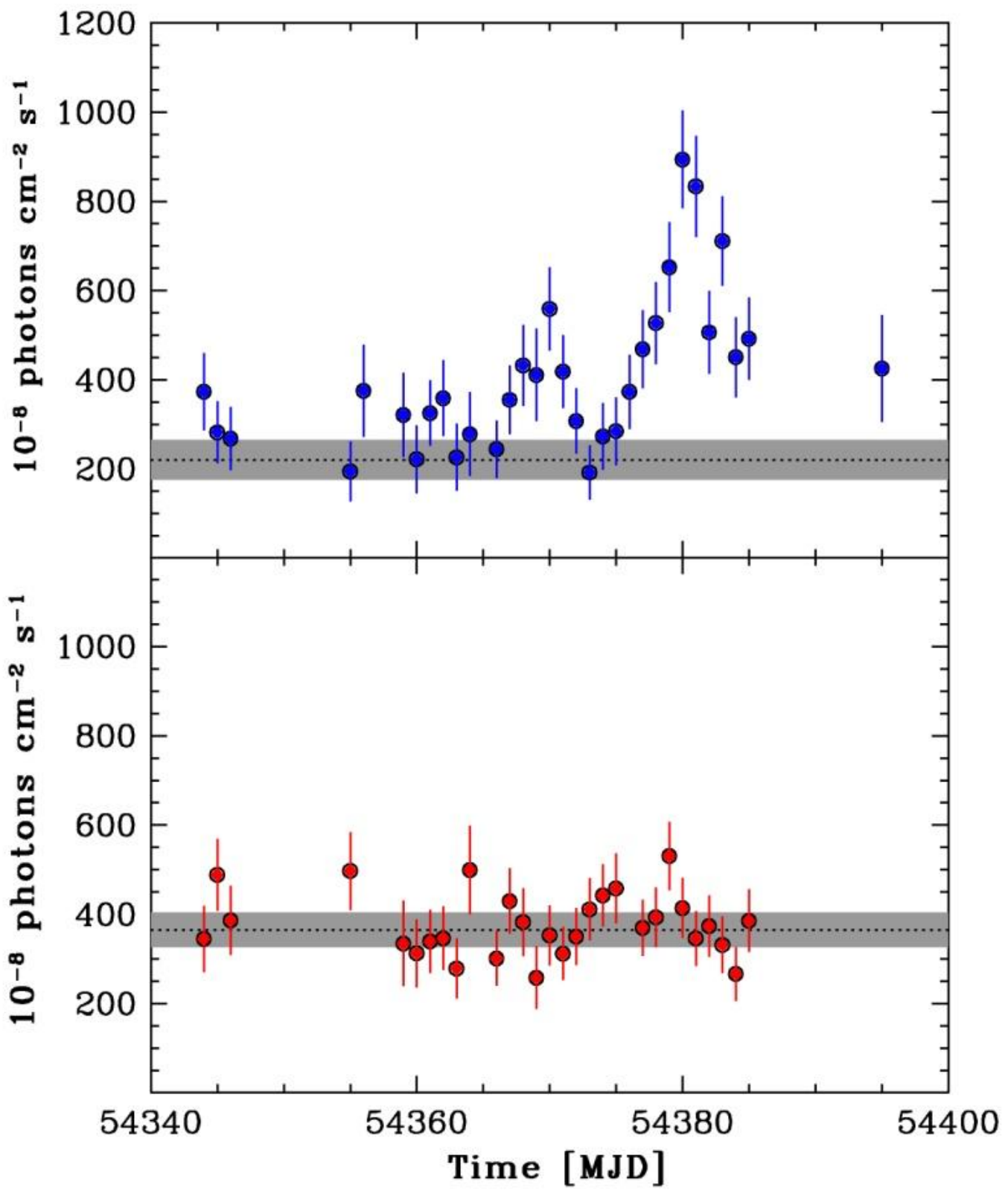

**Fig. S2** – Top panel: The AGILE gamma-ray light curve (1-day binning) of the Crab Pulsar/nebula above 100 MeV during the period 2007-08-28 – 2007-10-27 with the satellite in pointing mode. Bottom panel: same as the top panel light curve but for the nearby Geminga pulsar. Dashed lines and shadowed bands indicate the Crab average flux and the 3σ uncertainty range.

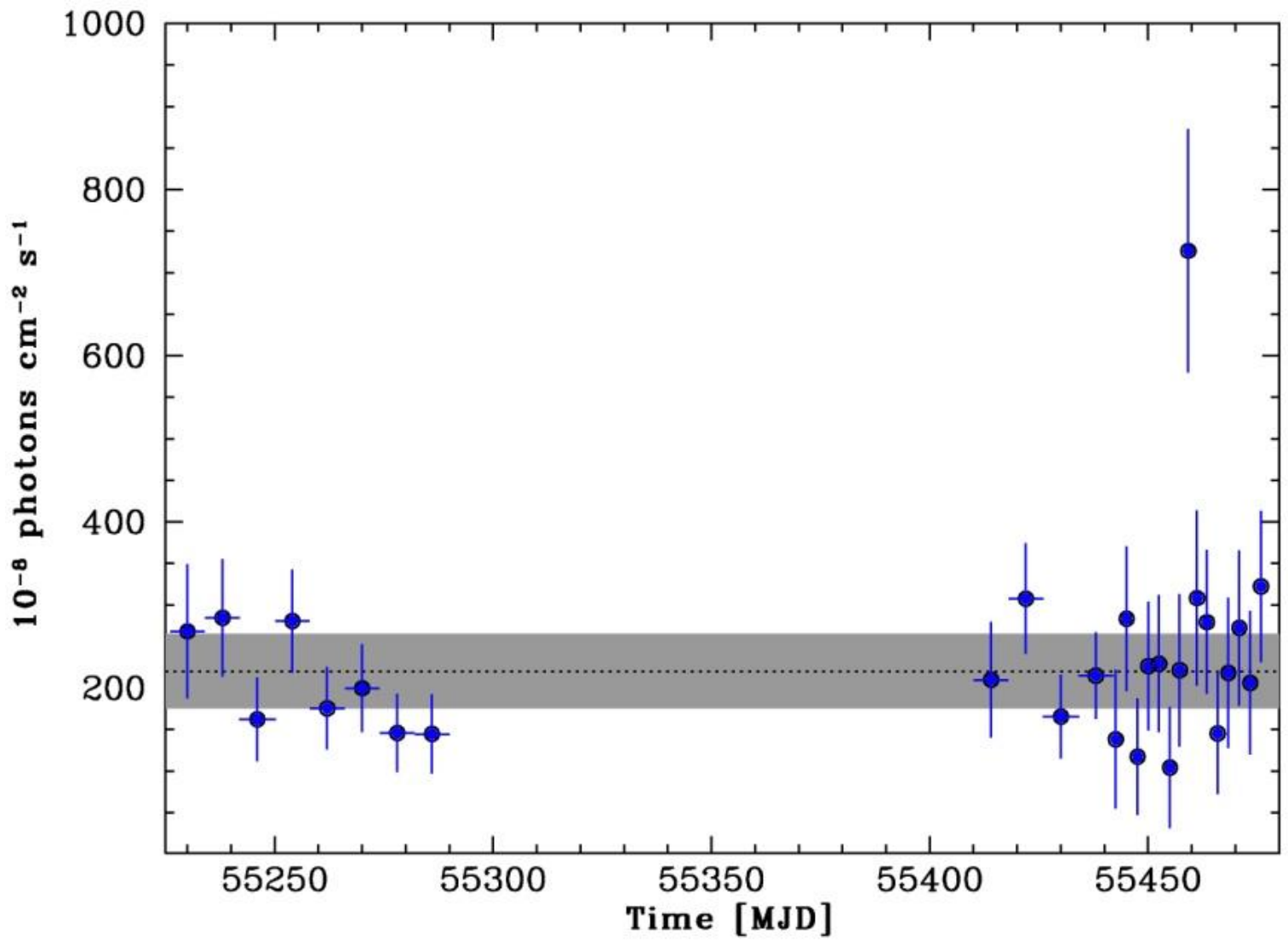

**Fig. S3** - The AGILE gamma-ray light curve of the Crab Pulsar/nebula above 100 MeV as observed with the satellite in spinning-mode. The light-curve covers the period from 2010-01-31 to 2010-10-07 . As for Fig. S1 and S2, the dashed line and shadowed band indicate the average Crab fluxes and the 3σ uncertainty range.

The gamma-ray data analysis was carried out with the standard AGILE/GRID FM3.119_2 filter, and with the MULTI-2 likelihood analysis package available at the AGILE Data Center. The Galactic diffuse gamma-ray emission was modeled using the AGILE standard model[17].

Fig. S1 shows the AGILE gamma-ray light-curve of the Crab Pulsar/Nebula (upper panel) and of the Geminga Pulsar (bottom panel), for a 1-day integration time, during the period 2007-09-01 – 2009-09-15, when AGILE was operating in pointing-mode. The dashed lines and the grey stripes indicate the average fluxes of the first AGILE-GRID catalogue[18] and the $3\sigma$ error, respectively.

The gamma-ray flare occurred between 2007 September 22 and 2007 October 12 is clearly visible in Fig. S1 and is shown in details in Fig. S2. In all remaining AGILE observations the Crab Pulsar/Nebula was detected with a flux compatible with its average one. The flux of the nearby Geminga pulsar was always observed as compatible with the value reported in the 1$^{st}$ AGILE catalogue, and we can thus exclude that the observed increase in the Crab flux during October, 2007 was due to a systematic effect.

Fig. S3 shows the AGILE light-curve (100 MeV – 5 GeV) in the period 2010-02-03 – 2010-10-07 collected with the satellite in spinning-mode. The September 2010 flare is shown in details in Fig. 1 of the main text.

In Fig. S1, as well as in Fig. S2 and in Fig. S3, the gamma-ray fluxes are calculated with a multi-source likelihood analysis taking into account the 2 most prominent gamma-ray sources within a radius of 15 degrees from the Crab Nebula and the Crab itself. The two other source are:

1AGL J0634+1748 (GEMINGA: F = (365 +/- 13) x $10^{-8}$ ph $cm^{-2}$ $s^{-1}$, $l$ =194.14, $b$ =4.36)

1AGL J0617+2236 (SNR IC443: F = (69 +/- 9) x $10^{-8}$ ph $cm^{-2}$ $s^{-1}$, $l$ =189.04, $b$ =3.07).

A maximum likelihood analysis determines that the Crab pulsar+Nebula has a persistent flux of F = (2.2 +/- 0.1) $\cdot 10^{-6}$ ph $cm^{-2}$ $s^{-1}$ for E > 100 MeV over the period from 2007-12-16 to 2009-10-15 (excluding the period of the October 2007 gamma-ray flare), at a significance of sqrt(TS) = 30.0. This result takes into account the diffuse gamma-ray background with Galactic and isotropic components, and is obtained considering all nearby sources with a fixed flux. An automated search procedure continuously performs a maximum likelihood analysis over maps of 2-day integrations. This algorithm (named "AGILE SPOT") is routinely used for the automated detection of gamma-ray flares from Galactic and extragalactic sources (see ref. 33). The SPOT algorithm is a twostep procedure that extracts excesses from counts maps and builds a list of candidate gamma-ray sources which are then analyzed by the likelihood method. For a candidate flaring source we require that the significance of the flux excess be larger than sqrt(TS) > 4.5. For a candidate flare from a persistent gamma-ray source (such as the Crab) we include in the null hypothesis the standard persistent flux and position of the candidate flaring source as well as the other sources in the nearby region. For the October 2010 Crab flare (of total flux $F_T$ = (7.2+/-1.4)$\cdot 10^{-6}$ $cm^{-2}$ $s^{-1}$) the excess flux at the Crab position is F' = (5.1+/-1.4) $\cdot 10^{-6}$ $cm^{-2}$ $s^{-1}$ which turns out to have a significance of sqrt(TS) = 4.8 over the persistent source of fixed flux F = (2.2 +/- 0.1) $\cdot 10^{-6}$ ph $cm^{-2}$ $s^{-1}$.

No significant pulsed flux and pulse profile variation is observed during and following the reported flare episodes.

## 4. SEARCH FOR A FIELD OBJECT IN THE CRAB NEBULA ERROR BOX

Extreme, short-term variability of the gamma-ray emission (E>100 MeV) from pulsar wind nebulae like the Crab nebula is so unusual and unexpected that we must take seriously the possibility that the emission comes not from the nebula itself but from a nearby source within the AGILE and FERMI error box, for example an unidentified blazar.

Blazars are the class of gamma-ray sources which more typically exhibit gamma-ray flares lasting a few days of the type we observe here. Knowing that the intrinsic distribution of blazars should be isotropic, we can use the 596 AGN associations above $|b| > 10°$ in the Fermi AGN Catalog[19] to estimate the probability of finding at least one Fermi-like blazar within the error box, which has a radius of[20] ~ 0.06°. The calculated probability is $\sim 1.6 \cdot 10^{-4}$, making it unlikely that the observed gamma-ray variability was due to an unidentified blazar.

As reported in the main text, several follow-up observations in the X-ray band showed the lack of any significant variation in the Crab pulsar flux, which suggests that the gamma-ray flare was likely associated with the Crab Nebula, although the contamination of other sources could not be ruled out. Moreover, by means of Swift observations, it has been possible to perform imaging in the soft X-ray band[21], which resulted in the lack of detection of new X-ray sources near the Crab. This was confirmed by the observations performed by *Chandra* and *HST*. Both X-ray and optical images showed an enhancement of the emission about 3 arcsec east of the pulsar with respect to previous observations performed by the two observatories.

## 5. Super-AGILE OBSERVATIONS

Since the Crab is a bright source in the hard X-ray band, we used the Super-AGILE 18-60 keV data in order to study if the flare detected in gamma-rays is accompanied by a variation in the pulsed emission. Given the higher Super-AGILE statistics for the source in the pointing operative mode, we concentrated on the 2007 flare. We corrected the photon arrival time to the Solar System Barycenter and we used the radio monthly ephemeris from the Jodrell Bank Centre for Astrophysics[22,23] to convert the time into rotational phase of the pulsar and produce "folded" light-curves.

We show in Fig. S4 the folded light-curves of the Crab pulsar accumulated during the flare phase, between 2007-09-27 12:15 UT and 2007-10-10 2007-10-13 10:53 UT (blue line), and during quiescence, from 2007-08-22 14:04 UT until 2007-09-24 04:20 UT (green line). The comparison of the two curves indicates that shape of the peaks in the light-curve is the same, without evidence of variation in the pulsar characteristics.

In order to search for structured variation in the Crab pulsar pulse-shape profile we performed a Runs Test of Randomness[24] on the differences between the two Super-AGILE folded light curves. We obtained a nearly-normal test statistic Z equal to 0.29 and a probability of obtaining a value of $Z \geq 0.29$ equal to 39%. This is consistent with null hypothesis of no difference between the light curves, and we can thus confirm that there are no changes in the two Super-AGILE hard X-ray pulse shape profiles.

## 6. Swift/XRT OBSERVATIONS

The AGILE detection of the 2010 gamma-ray flare triggered a ToO observation[25] of the Crab Nebula/Pulsar with the XRT instrument onboard the Swift satellite[26,27]. XRT is a focusing telescope with a CCD detector, and its band of sensitivity is roughly 0.5-10 keV. Starting from 2010 September 22 to September 28, Swift observed the Crab in 10 occasions, for a total exposure of about 17 ks, and the log of these observations is reported in Tab. S1.

XRT observations were performed in the Windowed Timing (WT) mode, that retains the full timing capability of the instrument: the central 200 CCD columns (corresponding to a field of view of 8 arcmin) are registered in one-dimensional mode, with a time resolution of 1.7 ms. Data were reduced accordingly to the standard procedures, barycenterized and folded using the Jodrell Bank ephemeris. For a comparison of the pulse shape in different source states, we reduced also two archival Crab observations, one performed in 2007 October 15 (just at the end of the 2007 gamma-ray flare) and in 2009 September 17 (in quiescence). The folded light-curves were normalized at the first peak height after subtracting the off-pulse counts (evaluated in the phase interval 0.60-0.83).

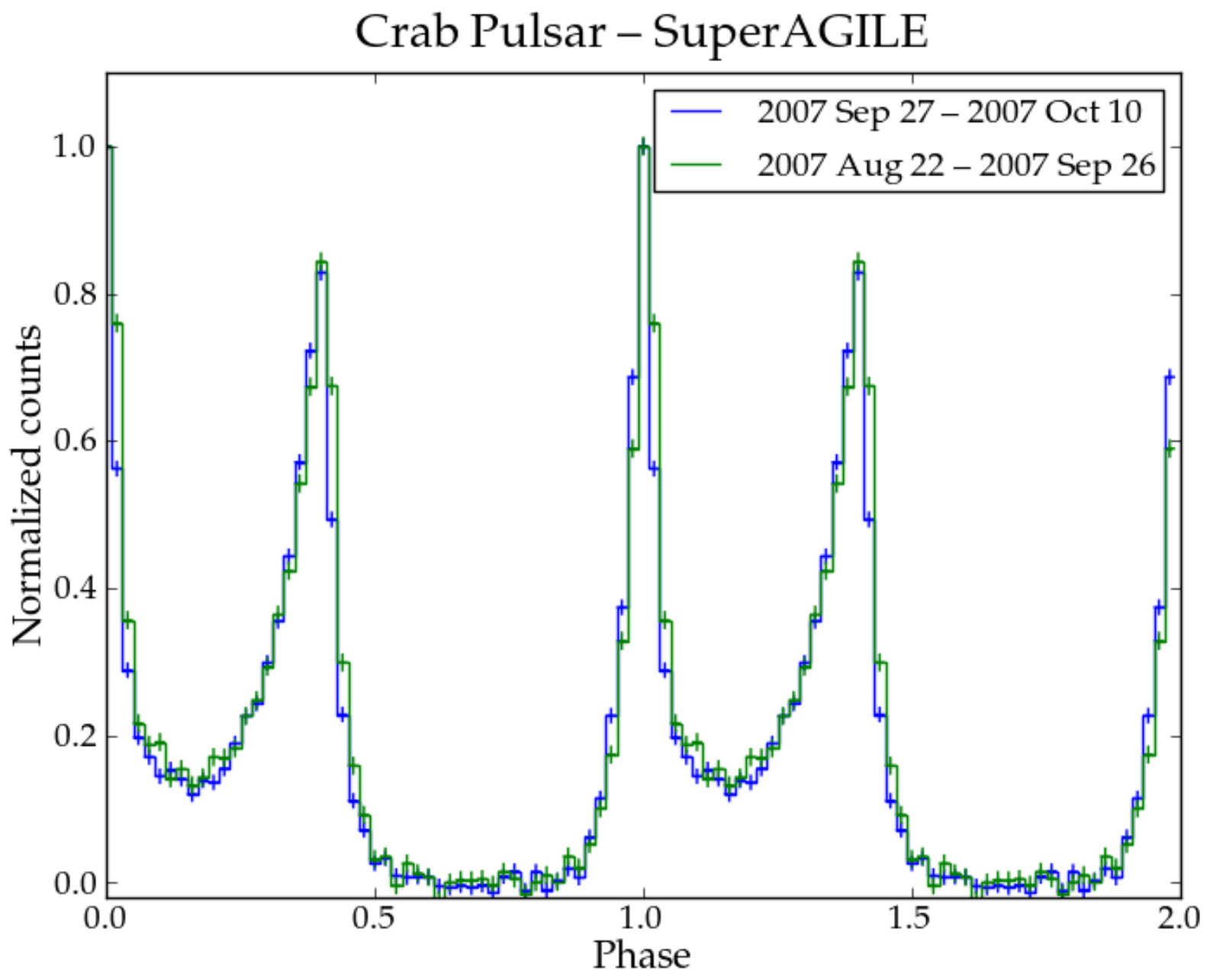

**Fig. S4** – Light-curves of the Crab pulsar as accumulated by SuperAGILE in the hard X-ray (18-50 keV) energy band during (blue line) and immediately before (green line) the flaring episode of October 2007 and converted in phase with the radio monthly ephemeris from the Jodrell Bank Centre for Astrophysics.

**Table S1** – Journal of the Swift/XRT observations of the Crab Nebula/Pulsar collected during the September, 2010 gamma-ray flare.

| Obs. ID | Date Start | Time Start (UT) | Exposure (s) |
|---|---|---|---|
| 00030371002 | 2010-09-22 | 16:40:11 | 1000 |
| 00030371003 | 2010-09-23 | 18:21:48 | 995 |
| 00030371004 | 2010-09-24 | 13:53:00 | 1040 |
| 00030371005 | 2010-09-24 | 01:02:17 | 1000 |
| 00030371006 | 2010-09-24 | 06:12:50 | 1000 |
| 00030371007 | 2010-09-24 | 20:03:28 | 1000 |
| 00030371008 | 2010-09-25 | 01:29:02 | 995 |
| 00030371012 | 2010-09-26 | 01:00:00 | 4385 |
| 00030371013 | 2010-09-27 | 01:26:00 | 5025 |
| 00030371014 | 2010-09-28 | 01:11:00 | 1920 |

We performed the Runs Test of Randomness[24] on the Swift/XRT light-curves. No significant pulse shape changes were detected between the September 2010 observations, and between the 2007, 2009 and 2010 observations (Fig. S5). Also a comparison of the Swift/XRT and AGILE/Super-AGILE light-curves with the Crab pulsed emission reported in literature[28,29,30] does not show significant modification in the pulse shape profiles during the two flaring episodes.

The lack of variations in the Crab soft and hard X-ray pulse shape profile collected during and/or immediately after the 2007 and the 2010 gamma-ray flares strengthen the confidence in the nebular origin for the Crab flaring activity. This agrees with the results of the pulsed gamma-ray light-curve analysis performed on the AGILE and Fermi[31] data and, also, with the analysis of the radio data[32].

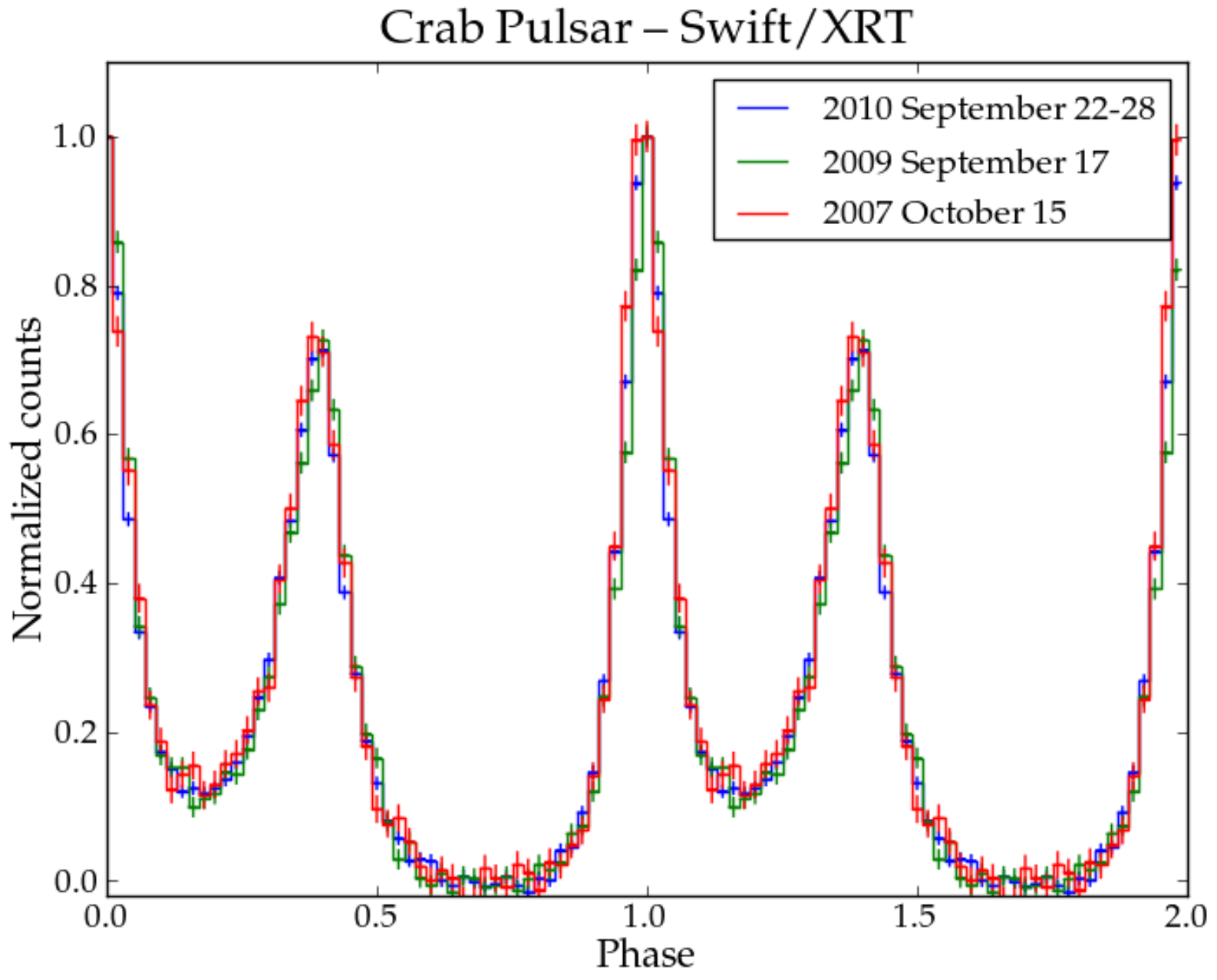

**Fig. S5** - Swift/XRT folded lightcurves (Window-timing mode, WT) accumulated in the energy band 0.5-10 keV. Red line: light curve accumulated during the 2007 flare ($T_{start}$: 2007-10-15 23:59 UT, exposure: 4485 s). Green line: reference light curve accumulated in September 2009 ($T_{start}$ 2009-09-17 13:26 UT, exposure: 6284 s). Blue line: light curve accumulated during (and after) the September 2010 flare (from 2010-09-22 16:40 UT to 2010-09-28 07:49 UT (10 observations), exposure: 17837.5 s). Phase and period were determined by using the radio monthly ephemeris from the Jodrell Bank Centre for Astrophysics. No significant variation in the pulse shape profile is present, strengthening the Nebular origin of the two gamma-ray flares reported in the text